\documentclass[onecolumn,preprint,showpacs,superscriptaddress]{revtex4-1}

\usepackage[utf8]{inputenc}
\usepackage{mathtools}
\usepackage{graphicx}
\usepackage{caption}
\usepackage{subcaption}
\usepackage{xcolor}
\usepackage{soul}
\usepackage{float}
\usepackage{amsmath}
\usepackage{amssymb}
\usepackage{amsfonts}
\usepackage{xcolor}
\usepackage{appendix}
\usepackage{orcidlink}

\begin{document}

\title{Bound and Scattering States in a Spacetime with Dual Topological Defects: Cosmic String and Global Monopole}

\author{L. G. Barbosa \orcidlink{0009-0007-3468-3718}}
\email{leonardo.barbosa@posgrad.ufsc.br}
\affiliation{Departamento de Física, CFM - Universidade Federal de \\ Santa Catarina; C.P. 476, CEP 88.040-900, Florianópolis, SC, Brazil}

\author{L. C. N. Santos \orcidlink{0000-0002-6129-1820}}
\email{luis.santos@ufsc.br}
\affiliation{Departamento de Física, CFM - Universidade Federal de \\ Santa Catarina; C.P. 476, CEP 88.040-900, Florianópolis, SC, Brazil}

\author{J. V. Zamperlini \orcidlink{0009-0002-9702-1555}}
\email{joao.zamperlini@posgrad.ufsc.br}
\affiliation{Departamento de Física, CFM - Universidade Federal de \\ Santa Catarina; C.P. 476, CEP 88.040-900, Florianópolis, SC, Brazil}

\author{F. M. da Silva \orcidlink{0000-0003-2568-2901}}
\email{franciele.m.s@ufsc.br}
\affiliation{Departamento de Física, CFM - Universidade Federal de \\ Santa Catarina; C.P. 476, CEP 88.040-900, Florianópolis, SC, Brazil}

\begin{abstract}
This paper explores the relativistic quantum motion of scalar bosons in the presence of mixed topological defects: cosmic strings and global monopoles. The Klein-Gordon equation with generalized Coulomb potentials is analyzed in this background. The effects of these topological defects on the equations of motion, phase shifts, and the S-matrix are examined in detail. Bound state solutions are derived from the poles of the S-matrix.  We provide analytical expressions for the energy spectrum of bound states, with particular attention to how the parameters of scalar and vector potentials affect the behavior of the system. 
Furthermore, we explore particular cases involving pure scalar, vector, and mixed scalar-vector potentials, showing how these scenarios impose particular conditions on the existence of bound states. Our results indicate that the solutions obtained associated with scattering and bound states depend significantly on the parameters of the topological defects.
\end{abstract}

\maketitle

\section{Introduction}
Topological defects, including cosmic strings and global monopoles, emerge as intriguing structures that may form during symmetry-breaking phase transitions in the early universe \cite{string12,string13,string14,string15,string16}. Cosmic strings are one-dimensional defects characterized by a linear mass density, capable of inducing significant gravitational effects. Their associated spacetime is locally flat yet globally curved, resulting in phenomena such as gravitational lensing and the gravitational Aharonov-Bohm effect. Conversely, global monopoles are zero-dimensional defects that exhibit a nontrivial vacuum structure, leading to a deficit solid angle in the spacetime. These topological features interact with quantum fields in significant ways, impacting the dynamics of particles situated in their vicinity. In the last years, the effect of cosmic strings on Schr\"{o}dinger \cite{wang2015exact,muniz2014landau,ikot2016solutions,ahmed2023effects}, Klein-Gordon \cite{santos2,string10,string9,string6,string5,string4,string2,neto2020scalar,boumali2014klein,ahmed2021effects,santos2018relativistic} and Dirac \cite{inercial10,inercial8,string8,string7,string1,hosseinpour2015scattering,marques2005exact,bakke2018dirac,hosseinpour2017scattering,cunha2020dirac,lima20192d} equations has been widely studied in the literature. In the same way, there have been several studies about the influence of global monopoles on non-relativistic \cite{global4,ahmed2024effects,alves2023approximate,alves2024exact,mustafa2023schrodinger,BezerradeMello:1996si}, bosonic \cite{global2,de2006exact,de2022klein,montigny2021exact,bragancca2020relativistic,ahmed2022relativistic} and fermionic \cite{global3,ali2022vacuum,ren1994fermions,bezerra2001physics} particles.

The study of quantum systems in curved spacetimes, including noninertial effects on physical systems \cite{inercial1,inercial2,inercial4,inercial5,inercial6,inercial7,inercial9,inercial10,inercial11,santos3,santos2019klein,santos2023non}, has led to significant advancements in the formulation of the Schrödinger, Klein-Gordon, and Dirac equations within curved geometries.  Notably, the behavior of hydrogen atoms in arbitrary curved spacetimes serves as a critical platform for understanding how spacetime curvature influences atomic energy levels. In this scenario, it has been demonstrated that the energy spectrum related to single-electron atoms in a curved spacetime diverges from that found in a typical flat Minkowski spacetime \cite{parker1}. Gravitational fields cause shifts in the energy levels, with curvature effects manifesting as perturbations in the relativistic fine structure. These shifts depend on the geometric characteristics of the defects and manifest differently across various energy levels, distinguishing them from typical gravitational redshifts or the Doppler effect.  Furthermore, Wheeler and Brill explore the behavior of neutrinos within a curved metric in \cite{neutrino1}, offering an in-depth examination of how gravitational fields interact with neutrinos. Previous studies have explored the quantum dynamics of particles in the presence of topological defects, including their emission from black holes with such defects \cite{jusufi2015scalar}. Additionally, scalar bosons with Coulomb potentials in the background of cosmic strings have been analyzed, revealing significant effects on scattering processes and bound state solutions \cite{neto2020scalar}. 
In a recent study \cite{string11}, the solutions of the Klein-Gordon equation were analyzed within G{\"o}del-type spaces containing an embedded cosmic string. It was demonstrated that the presence of this cosmic string in the spacetime interferes with the degeneracy of energy levels in Som-Raychaudhuri, spherical, and hyperbolic G{\"o}del spaces. Additionally, the influence of magnetic fields on the spacetime metric was explored in \cite{santos1}. These works underscore the importance of understanding how topological defects influence quantum systems, especially with regard to their effects on particle dynamics and energy levels.
 
While prior studies have explored the effects of individual topological defects on quantum systems, including the modifications of energy levels caused by cosmic strings and global monopoles, the combined influence of these defects on quantum states has not been thoroughly examined in a unified theoretical framework.
In this paper, we investigate the relativistic quantum motion of scalar bosons in a spacetime characterized by mixed topological defects: cosmic strings and global monopoles. We analyze the Klein-Gordon equation with generalized Coulomb potentials within this combined background, focusing on how these topological structures modify the equations of motion, phase shifts, and the S-matrix. \textcolor{black}{Our main motivation is to extend previous investigations that only consider one of the possible potentials in the Klein-Gordon equation to be a Coulomb-like potential and consider only a single topological defect as the background. This way, we can analyze the behaviors that arise from these combined effects and, additionally, recover previous results as special cases of our solutions.}
Vector $(p^\nu)$ and scalar $(M)$ potentials are taken into account through the replacements of momentum operator
\begin{align}    
    M &\rightarrow M + V_s, \\
    p^\nu & \rightarrow p^\nu + eA^\nu,
\end{align}
where $V_s$ behaves as a scalar under a Lorentz transformation, $e$ is a real coupling parameter and $A^{\nu}$ as a vector quantity associated with the electromagnetic field. Here, we consider an additional interaction represented by the replacement
\begin{equation}
    p^{\nu}p_\nu \rightarrow (p^{\nu}-iX^\nu)(p_{\nu} + iX_\nu),
\end{equation}
where $X^{\nu}$ 
is a not minimally coupled vector. 
By deriving bound state solutions from the poles of the S-matrix, we elucidate the dependencies of these solutions on the parameters associated with both the cosmic string and global monopole.

The remainder of this paper is organized as follows: In Section \ref{sec2}, we describe the topological defects under consideration, namely the cosmic string and global monopole, and introduce the generalized metric used to account for their combined effects on spacetime. Section \ref{sec3}, presents the derivation of the curved wave  equation and discusses the solutions of the Klein-Gordon equation in this curved background. In Section \ref{sec4}, we focus on the generalized Coulomb potential, analyzing its impact on both bound and scattering states. Particular cases of the scalar and vector potentials are discussed in Section \ref{sec5}, where we explore the conditions for the existence of bound states \textcolor{black}{and non-relativistic limits}. Finally, in Section \ref{sec6}, we provide a detailed discussion of the results, emphasizing the role of the topological defects on the quantum dynamics of particles and concluding with suggestions for future research directions.

\textcolor{black}{Throughout the development of the paper we employ Planck units ($G=c=\hbar=1$), which renders all quantities dimensionless.}

\section{Topological Defects and Generalized Metric} \label{sec2}

As can be seen from various works in the literature, topological defects, such as cosmic strings and global monopoles, lead to interesting modifications in the spacetime geometry. Thus, the combination of the two defects can lead to new features of the spacetime and, consequently, imply new effects on the physical systems described in this geometry. In this section, we consider a generalized spacetime metric that incorporates the effects of both defects, taking into account their associated deficit angles. 

The metric describing the spacetime in the presence of both a cosmic string and a global monopole in spherical coordinates is given by \cite{bezerra2002bremsstrahlung,jusufi2015scalar}:
\begin{equation}
    ds^{2} = -dt^{2} + dr^{2} + \beta^{2} r^{2} d\theta^{2} + \alpha^{2} \beta^{2} r^{2} \sin^{2}\theta d\phi^{2},
\end{equation}
where $ -\infty < t<  \infty$, $0 \leq r < \infty$, $0 \leq \theta \leq \pi/2 $, and $0 \leq \phi < 2\pi $. The quantities \(\alpha\) and \(\beta\) represent the parameters related to the cosmic string and global monopole, respectively. These parameters are defined as follows: The parameter \(\beta \in (0,1]\), which is associated with the global monopole, accounts for the deficit solid angle and is given by:
\begin{equation}
    \beta^{2} = 1 - 8\pi \tilde{\eta}^{2},
\end{equation}
where \(\tilde{\eta}\) is the symmetry-breaking scale related to the formation of the monopole. The strength of the monopole's gravitational field depends on \(\tilde{\eta}\), with larger values leading to a more pronounced solid angle deficit.

The parameter \(\alpha \in (0,1]\), associated with the cosmic string describes the deficit angle introduced by the string and can be written as
\begin{equation}
    \alpha^{2} = \left(1 - 4\tilde{\mu}\right)^{2},
\end{equation}
where \(\tilde{\mu}\) represents the linear mass density of the cosmic string. The conical structure of the spacetime is influenced by \(\tilde{\mu}\), with \(\alpha\) approaching unity in the absence of the string's gravitational effects.
Together, \(\alpha\) and \(\beta\) modify the angular components of the metric, reflecting the topological characteristics of both the cosmic string and the global monopole.
\textcolor{black}{ Note that in astrophysical problems it is common to study the cases $\alpha <1$ and $\beta <1$. On the other hand, in condensed matter applications it is possible to exploit the well-known analogy between disinclination in solids and cosmic strings. In solids, it is considered a geometry originates from the metric associated to the spacial sector of astrophysical spacetime metric \cite{nelson2002defects,kibble2008introduction}.  In these systems, the cases $\alpha >1$ and $\beta >1$ can be considered in several applications. For instance, the values corresponding to $\alpha = 1$ and $\beta >1$ were analyzed in  \cite{Furtado:1994nq} where the authors concluded that negative declinations act as attractors to  electrons and holes giving rise to bound states in a material.}

\section{Field Equations and Solutions} \label{sec3}

To analyze the quantum dynamics of scalar bosons in the presence of topological defects, we start by considering the following generalized Klein-Gordon equation. This equation incorporates scalar and vector potentials, along with nonminimal couplings, and is written as:
\begin{equation}
    -\frac{1}{\sqrt{-g}}D_{\mu}^{\left(+\right)}\left(g^{\mu\nu}\sqrt{-g}D_{\nu}^{\left(-\right)}\psi\right)+\left(M+V_{s}\right)^{2}\psi=0,
    \label{3.1}
\end{equation}
where $\psi=\psi(t,r,\theta,\phi)$ represents the bosonic field, $g=\det(g_{\mu\nu})$ and the covariant derivative terms in this expression, \(D_{\mu}^{\left(\pm\right)}\), are defined as:
\begin{equation}    D_{\mu}^{\left(\pm\right)}=\partial_{\mu}\pm X_{\mu}-ieA_{\mu},
\label{3.2}
\end{equation}
where the scalar $(V_s)$ and vector \((A_{\mu}\) and \(X_{\mu})\) potentials are functions of the radial coordinate only\textcolor{black}{, and $e$ is the associated electric charge}. The potential \(A_{\mu}\) describes the electromagnetic interaction and is given by:
\begin{equation}    A_{\mu}=\left(A_{t}\left(r\right),0,0,0\right).
\label{3.3}
\end{equation}
Meanwhile, \(X_{\mu}\), the nonminimal vector potential, has both temporal and radial components:
\begin{equation}    X_{\mu}=\left(X_{t}\left(r\right),X_{r}\left(r\right),0,0\right),
\label{3.4}
\end{equation}
and the scalar potential, $V_s$, is coupled to the mass of particle $M$. It is worth noting that $V_s$ and $X_{\mu}$ do not couple to the charge of the particle, in contrast to $A_{\mu}$.   At this point, we look for solutions to the Klein-Gordon equation considering that the wave function can be written in the form:
\begin{equation}   \psi\left(t,r,\theta,\phi\right)=\frac{u\left(r\right)}{r}f\left(\theta\right)e^{im\phi}e^{-i\varepsilon t},
\label{3.5}
\end{equation}
where $m=\pm 1,\pm 2,\pm 3,...$ and $\varepsilon$ is the energy of the particle. By substituting Eq. (\ref{3.5}) into expression (\ref{3.1}), we can see that the angular function \(f(\theta)\) satisfies the following differential equation, which includes the deficit angle parameter \(\alpha\) associated with the topological defects:  
\begin{equation}
  \left[\frac{1}{\sin \theta}\frac{d}{d\theta}\left(\sin \theta\frac{d}{d\theta}\right)+\left(\lambda_{\alpha}-\frac{m^{2}}{\alpha^{2}\sin^2 \theta}\right)\right]f\left(\theta\right)=0.
  \label{3.6}
\end{equation}

With \(\lambda_{\alpha}\) being related to the eigenvalues of the angular momentum operator in the presence of the defect, given by
\begin{equation}    \lambda_{\alpha}=l_{\alpha}\left(l_{\alpha}+1\right) \quad,
\label{3.7}
\end{equation}
where $l_\alpha = n + |m_\alpha| = l + |m|(1/\alpha - 1)$ with $n \in \mathbb{N}$, $l = n + |m|$, and $m_\alpha = m/\alpha$, ranging within $-l_\alpha \leq m_\alpha \leq l_\alpha$. Here, $l$ and $m$ denote the orbital angular momentum and magnetic quantum numbers in flat space, respectively.

For the radial part, \(u(r)\), we arrive at a differential equation which governs the behavior of the wave function in the radial direction, including contributions from both topological defect parameters. The radial equation obtained is written as
\begin{equation}
    \frac{d^{2}u\left(r\right)}{dr^{2}}+\left(K^{2}-V_{\mathrm{eff}}^2-\frac{l_{\alpha}\left(l_{\alpha}+1\right)}{\beta^{2}r^{2}}\right)u\left(r\right)=0 \quad,
    \label{3.8}
\end{equation}
where the term \(K^{2}\) is defined as
\begin{equation}
    K^{2}=\varepsilon^{2}-M^{2} \quad.
\end{equation}

Finally, the effective potential \(V_{\mathrm{eff}}\) takes into account the contributions from the scalar and vector potentials, as well as the nonminimal interaction terms:
\begin{align}
V_{\mathrm{eff}}^2 = & \, V_{s}^{2} - e^{2}A_{t}^{2} + 2\left(MV_{s} - e\varepsilon A_{t}\right) \notag \\
& + \frac{\partial X_{r}}{\partial r} + \frac{2}{r}X_{r} + X_{r}^{2} - X_{t}^{2}\label{3.9} \quad.
\end{align}

The effective potential \(V_{\mathrm{eff}}\) encapsulates the influence of the external interactions, including both scalar and vector potentials, as well as the contributions from the topological defects present in the system. This structure provides a framework for analyzing the dynamics of scalar bosons in the combined cosmic string and global monopole background.

\section{The Generalized Coulomb Potential} \label{sec4}

We introduce a generalized Coulomb potential by considering the following forms for the scalar and vector potentials:
\begin{equation}
    A_{t} = \frac{\gamma_{t}}{r}, \quad V_{s} = \frac{\gamma_{s}}{r}, \quad X_{t} = \frac{\delta_{t}}{r}, \quad X_{r} = \frac{\delta_{r}}{r},
\end{equation}
where \(\gamma_t\), \(\gamma_s\), \(\delta_t\), and \(\delta_r\) are constants associated with the potentials in the system. Defining the parameters:
\begin{align}
     \alpha_{1} = 2\left(M\gamma_{s} - e\varepsilon\gamma_{t}\right),
\end{align}
\begin{align}
     \alpha_{2} = \delta_{r}\left(\delta_{r}+1\right) + \gamma_{s}^{2} - e^{2}\gamma_{t}^{2} - \delta_{t}^{2},
\end{align}
the radial equation takes the form:
\begin{equation}
    \frac{d^{2}u\left(r\right)}{dr^{2}} + \left(K^{2} - \frac{\alpha_{1}}{r} - \frac{\alpha_{2} + \frac{l_{\alpha}\left(l_{\alpha}+1\right)}{\beta^{2}}}{r^{2}}\right)u\left(r\right) = 0.
    \label{sch}
\end{equation}

This equation can be seen as a Schrödinger-type differential equation with a Coulomb-like potential. For this effective potential to have a well-defined structure, the condition $\alpha_1 < 0$ is necessary. In addition, bound states satisfy the relation $|\varepsilon| < M$. In this way, these conditions establish the possible values for the energy spectrum of the bound states.
Equation (\ref{sch}) can be simplified by introducing the variable change \(z = -2iKr\). The equation then becomes:
\begin{equation}
    \frac{d^{2}u}{dz^{2}} + \left(-\frac{1}{4} - i\frac{\alpha_{1}}{2Kz} - \frac{\alpha_{2} + \frac{l_{\alpha}\left(l_{\alpha}+1\right)}{\beta^{2}}}{z^{2}}\right)u = 0 \quad.
\end{equation}

To further simplify, we define the following parameters:
\begin{equation}\label{Eta_Gamma}
    \eta = \frac{\alpha_{1}}{2K} \quad, \qquad \gamma_{l}^{2} = \frac{1}{4} + \frac{l_{\alpha}\left(l_{\alpha}+1\right)}{\beta^{2}} + \alpha_{2} \quad.
\end{equation}

With these definitions, the radial equation takes the form of Whittaker's equation:
\begin{equation}
    \frac{d^{2}u\left(z\right)}{dz^{2}} + \left(-\frac{1}{4} - \frac{i\eta}{z} + \frac{\frac{1}{4} - \gamma_{l}^{2}}{z^{2}}\right)u\left(z\right) = 0 \quad.
\end{equation}

The well-defined solution to this equation near the origin is given by:
\begin{equation}
    u\left(z\right) = A e^{-\frac{z}{2}} z^{\frac{1}{2}+\gamma_{l}} M\left(\frac{1}{2}+\gamma_{l}+i\eta, 1+2\gamma_{l}, z\right),
    \label{hiper}
\end{equation}
where \(M(a,b;z)\) is the confluent hypergeometric function and $A$ a constant. For large \(|z|\), the asymptotic behavior of \(M(a,b;z)\) is approximated by:
\begin{align}
M\left(a,b;z\right) \simeq & \, \frac{\Gamma(b)}{\Gamma(b-a)} e^{-\frac{i}{2}\pi a} |z|^{-a} \notag \\
& + \frac{\Gamma(b)}{\Gamma(a)} e^{-i\left(|z| + \frac{\pi}{2}(a-b)\right)} |z|^{a-b}.
\end{align}

This asymptotic expansion allows us to understand the behavior of the wave function at large distances, which is of interest when studying scattering processes.

\subsection{Scattering States}

For the analysis of scattering states, we begin by considering the asymptotic regime where \(|z| \gg 1\) and \(K \in \mathbb{R}\). In this limit, the radial solution behaves as:
\begin{equation}
    u\left(r\right) \simeq \sin\left(Kr - \frac{l\pi}{2} + \delta_{l}\right) \quad,
\end{equation}
where the phase shift \(\delta_{l}\) is given by:

\begin{equation}
   \delta_{l}=\frac{\pi}{2}\left(l+\frac{1}{2}-\gamma_{l}\right)+\mathrm{arg}\Gamma\left(\frac{1}{2}+\gamma_{l}+i\eta\right) \quad.
\end{equation} 

For scattering in spherically symmetric potentials, the scattering amplitude can be expressed as a partial wave series:
\begin{equation}
    f\left(\theta\right) = \sum_{l=0}^{\infty} \left(2l+1\right) f_{l} P_{l}\left(\cos \theta \right) \quad,
\end{equation}
where \(P_{l}\left(\cos \theta\right)\) are the Legendre polynomials and \(f_{l}\) represents the contribution of each partial wave. 
Considering the S-matrix, \(S_{l}\), for each angular momentum quantum number \(l\), we have:
\begin{equation}
    S_{l} = e^{2i\delta_{l}} = e^{i\pi\left(l+\frac{1}{2}-\gamma_{l}\right)} \frac{\Gamma\left(\frac{1}{2} + \gamma_{l} + i\eta\right)}{\Gamma\left(\frac{1}{2} + \gamma_{l} - i\eta\right)} \quad.
\end{equation}

This expression for \(S_{l}\) encapsulates the phase shifts induced by the generalized Coulomb potential and the topological defects, providing key information about the scattering process.

\textcolor{black}{An important regime to be considered is the non-relativistic one. Evaluating the radial equation (\ref{sch}), we note that by taking
\begin{equation}
    \frac{l_{\alpha}\left(l_{\alpha}+1\right)}{\beta^{2}} \gg \alpha_{2},
\end{equation}
the radial Schrödinger equation is recovered \cite{BezerradeMello:1996si}. Consequently, setting $\alpha_2 \approx 0$ implies that
\begin{equation}
    \gamma_{l} \approx \sqrt{\frac{1}{4}+\frac{l_{\alpha}\left(l_{\alpha}+1\right)}{\beta^{2}}}.
\end{equation}}

\textcolor{black}{By setting the cosmic string parameter $\alpha=1$, the result obtained by Bezerra-Furtado \cite{BezerradeMello:1996si} is recovered. In addition, by setting the global monopole parameter $\beta=1$, one obtains $\gamma_l \approx l+\frac{1}{2}$, so that
\begin{equation}
    \delta_{l} \approx \operatorname{arg}\Gamma\left(1+l+i\eta\right).
\end{equation}}

\textcolor{black}{Recalling the definition of $\eta$ in equation (\ref{Eta_Gamma}) and taking $\gamma_t=0$, while expressing the energy in terms of the non-relativistic energy $E_{\text{nr}}$ and the rest mass $M$ \cite{greiner2000relativistic},
\begin{equation}\label{Approx}
    \varepsilon = M + E_{\text{nr}}, \quad E_{\text{nr}} \ll M,
\end{equation}
we can write
\begin{equation}
    \eta \approx \sqrt{\frac{M}{2E_{\text{nr}}}}\,\gamma_{s}.
\end{equation}}

\textcolor{black}{It is noted that as $E_{\text{nr}} \rightarrow 0$, $\eta \rightarrow \infty$, which motivates the treatment of the asymptotic expansion of the argument. In this limit, the phase shift can be written as
\begin{equation}
    \delta_{l} \approx \eta\ln\left(\eta\right) - \eta + \frac{\pi}{2}\left(l+\frac{1}{2}\right),
\end{equation}
and correspondingly the S-matrix is given by
\begin{equation}
    S_{l} \approx e^{i\pi\left(l+\frac{1}{2}\right)} e^{2i\left[\eta\ln\left(\eta\right) - \eta\right]}.
\end{equation}
This result is consistent with the expected outcome for flat space in the non-relativistic regime (with $\delta_t = \delta_r = \gamma_t = 0$), thereby demonstrating consistency with the relevant limits and the literature.}

\subsection{Bound States}
We can obtain solutions associated with bound states from the poles of the S-matrix. In this case, $K$ becomes imaginary and we can consider $K \rightarrow i|K|$. The S-matrix becomes infinite under the condition:
\begin{equation}
    \frac{1}{2} + \gamma_{l} + i\eta = -N,
\end{equation}
where \(N = 0, 1, 2, ..\). Defining:
\begin{equation}
    \mu = N + \frac{1}{2} + \gamma_{l},
\end{equation}
and substituting the expression for \(\eta\), the following quadratic equation for \(\varepsilon\) is obtained:
\begin{equation}\label{Eq_Espect}
    \left(e^{2}\gamma_{t}^{2} + \mu^{2}\right)\varepsilon^{2} - 2M\gamma_{s}e\gamma_{t}\varepsilon + \left(\gamma_{s}^{2} - \mu^{2}\right)M^{2} = 0,
\end{equation}
which can be solved to yield the energy levels for the bound states:
\begin{equation}
    \varepsilon_{\pm} = \frac{M}{\left(1 + \frac{e^{2}\gamma_{t}^{2}}{\mu^{2}}\right)} \left( \frac{\gamma_{s}}{\mu} \frac{e\gamma_{t}}{\mu} \pm \sqrt{1 + \frac{e^{2}\gamma_{t}^{2}}{\mu^{2}} - \frac{\gamma_{s}^{2}}{\mu^{2}}} \right).
    \label{energia}
\end{equation}

\begin{figure}[H]
    \centering
    \includegraphics[scale=0.5]{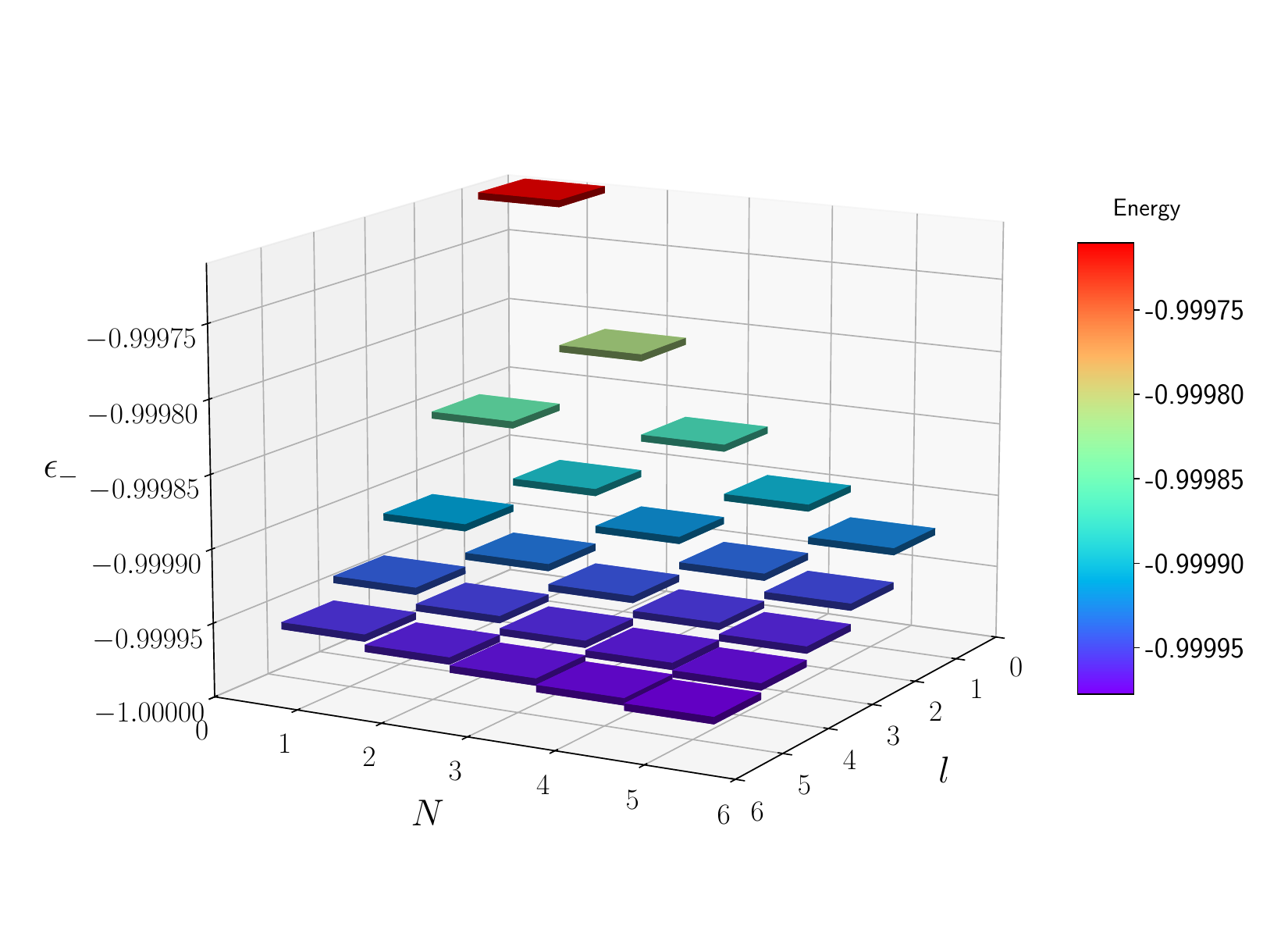}
    \caption{Plot of the negative energy spectrum dependent on the quantum numbers $N$ and $l$, which can take the values \{1,2,3,4,5\}. The parameters set for this plot were $\delta_r = 0.5$, $\delta_t = 0.6$, $\gamma_s = 0.7$, $e\gamma_t = -0.8$, $\alpha = 0.9$, $m = 1$, $M=1$, and $\beta=0.6$.}
    \label{fig:EnergySpectraxNxl-Beta0.6}
\end{figure}

\begin{figure}[H]
    \centering
    \includegraphics[scale=0.5]{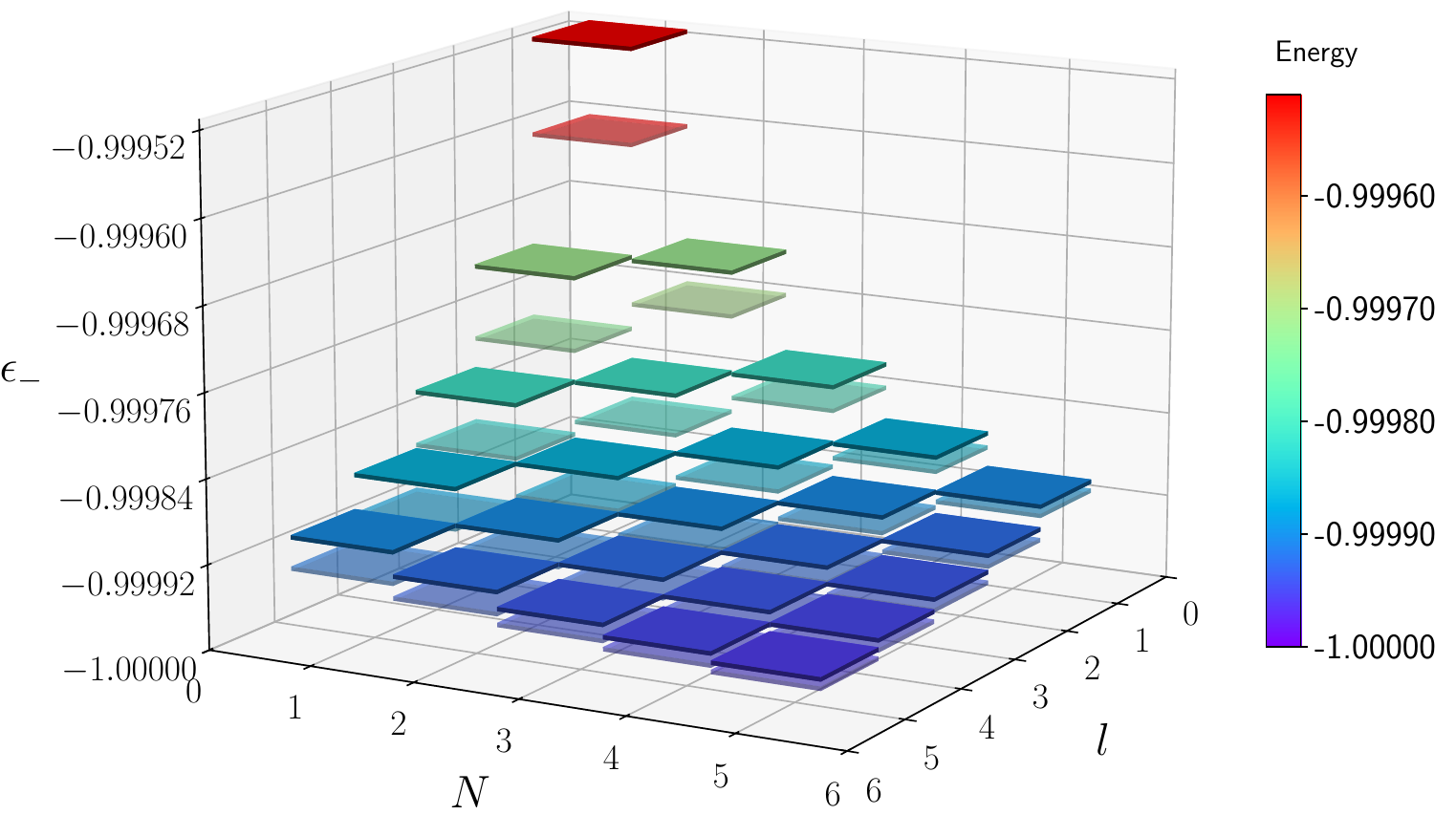}
    \caption{Plot of the negative energy spectrum dependent on the quantum numbers $N$ and $l$, which can take the values \{1,2,3,4,5\}, for $\beta=1$ (solid tiles) and $\beta=0.8$ (semi-transparent tiles), with parameters $\delta_r = 0.5$, $\delta_t = 0.6$, $\gamma_s = 0.7$, $e\gamma_t = -0.8$, $\alpha = 0.9$, $m = 1$, and $M=1$.}
    \label{fig:EnergySpectraBeta0.8And1}
\end{figure}

This result provides the energy spectrum of the bound states, which depends on the parameters of the generalized Coulomb potential and the topological defects. The expression highlights the role of the constants \(\gamma_{t}\), \(\gamma_{s}\), and \(\mu\), which characterize the potentials and the orbital angular momentum quantum numbers. \textcolor{black}{\autoref{fig:EnergySpectraxNxl-Beta0.6} shows the energy levels as function of the quantum numbers $N$ and $l$ for the case which $\beta=0.6$. Moreover, \autoref{fig:EnergySpectraBeta0.8And1} helps understand the effect of $\beta$ on the energy levels, if $\beta=1$ we recover the result from \cite{neto2020scalar} and then we can clearly see the effect of the global monopole parameter on the energy levels: as $\beta$ diminishes from the previous case ($\beta=1$) the energy levels decrease towards the asymptotic value of $-M$ as the quantum numbers $N$ and $l$ increase in both cases.}

\begin{figure}[H]
    \centering
    \includegraphics[scale=0.5]{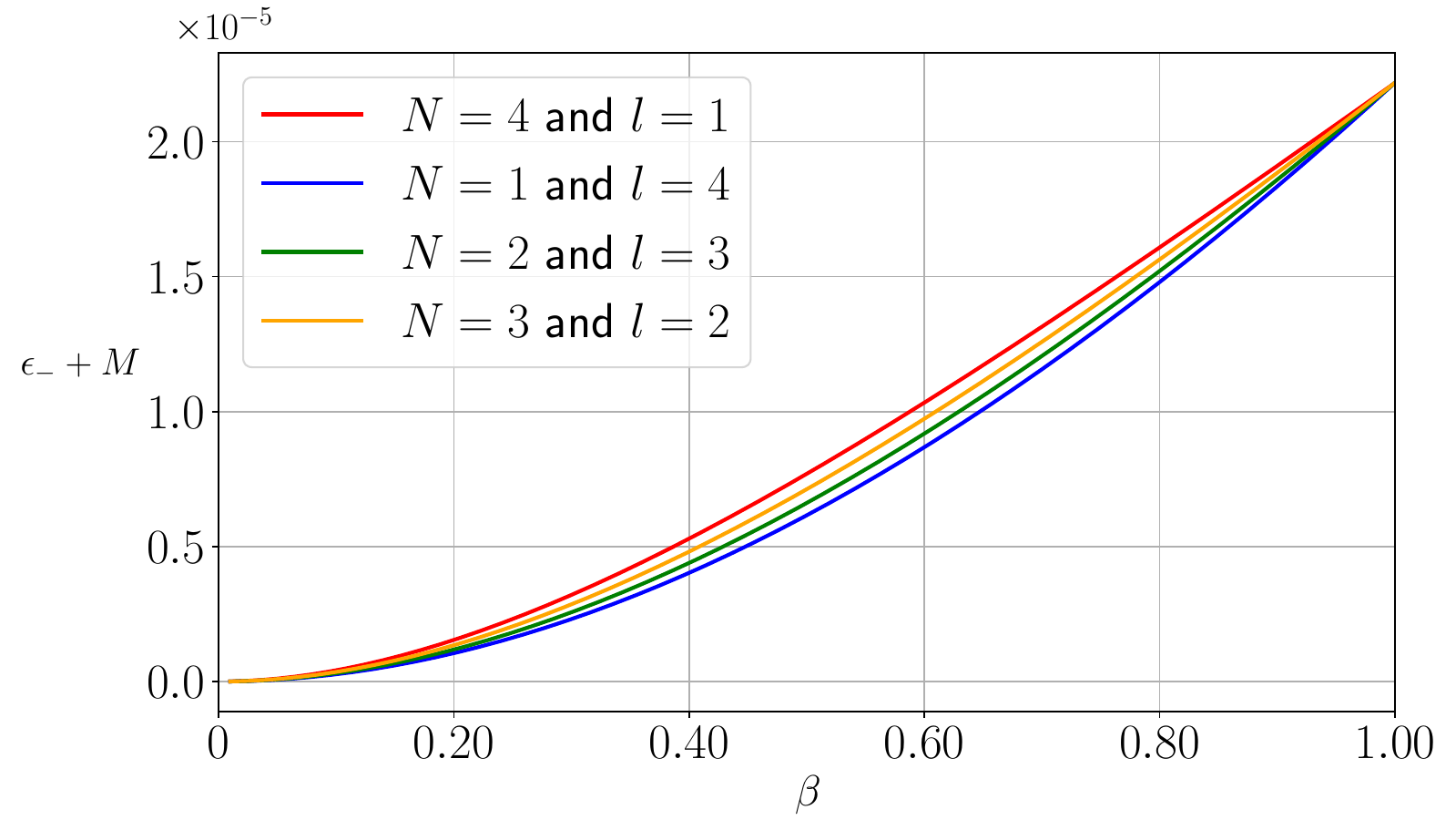}
    \caption{Plots of selected bound state energy levels plus $M$ which $N+l=5$ against $\beta$, with parameters $\delta_r = 0.5$, $\delta_t = 0.6$, $\gamma_s = 0.7$, $e\gamma_t = -0.8$, $\alpha = 0.1$, $m = 1$, and $M=1$.}
    \label{fig:energyAgainstBeta}
\end{figure}

\textcolor{black}{Another feature is the varying shift of quantized energy levels for distinct $\{N,l\}$ combinations, which is affected by the angular momentum quantum number at moderate $\beta$ values. Within flat spacetime and cosmic string contexts, energy levels are nearly degenerate for arrangements sharing the same $N+l$ sum. Conversely, when a global monopole parameter is introduced, configurations with higher $l$ demonstrate a noticeably greater decrease in energy, especially at intermediate $\beta$ values, as illustrated in Figures \ref{fig:EnergySpectraBeta0.8And1} and \ref{fig:energyAgainstBeta}, whereas in the later we add $M$ to the energy for better visualization.}

\begin{figure}[H]
    \centering
    \includegraphics[scale=0.5]{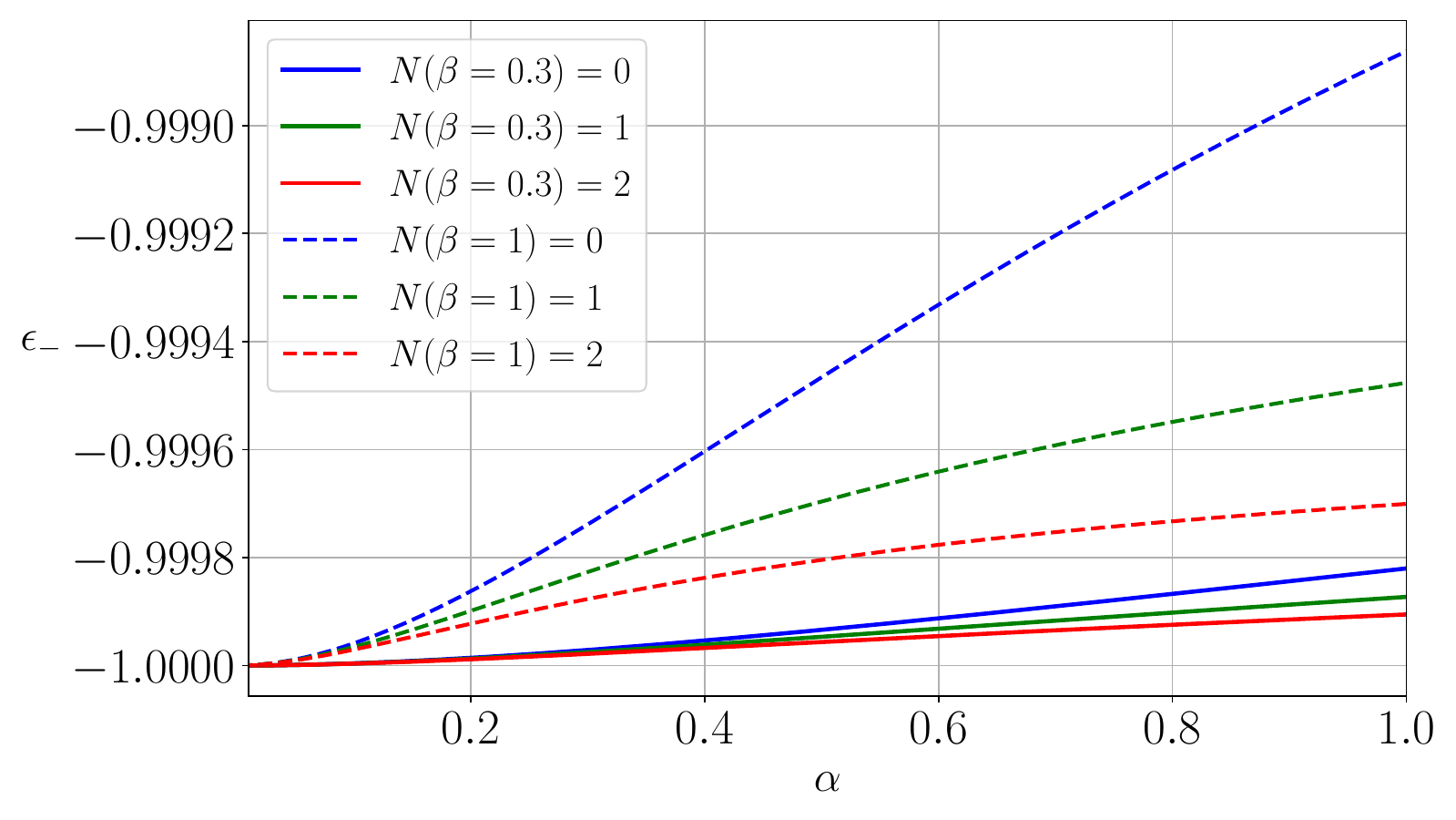}
    \caption{Plots of energy against angular deficit for three values of $N$, presenting the case devoid of global monopole ($\beta=1$) depicted with dashed lines, while the presence of a global monopole given by $\beta=0.3$ is shown using solid lines, with parameters $\delta_r = 0.5$, $\delta_t = 0.6$, $\gamma_s = 0.7$, $e\gamma_t = -0.8$, $\alpha = 0.9$, $m = 1$, and $M=1$.}
    \label{fig:EnergyLevelsMinusVsAlphaBeta0.3And1}
\end{figure}

\textcolor{black}{Similarly, still comparing with \cite{neto2020scalar}, we can see from \autoref{fig:EnergyLevelsMinusVsAlphaBeta0.3And1} that the behavior of the energy levels relative to the angular deficit is the same, increasing as $\alpha$ grows, as expected, but as $\beta$ decreases in value, so does all the energy levels, strengthening the bond.}

\begin{figure}[H]
    \centering
    \includegraphics[scale=0.5]{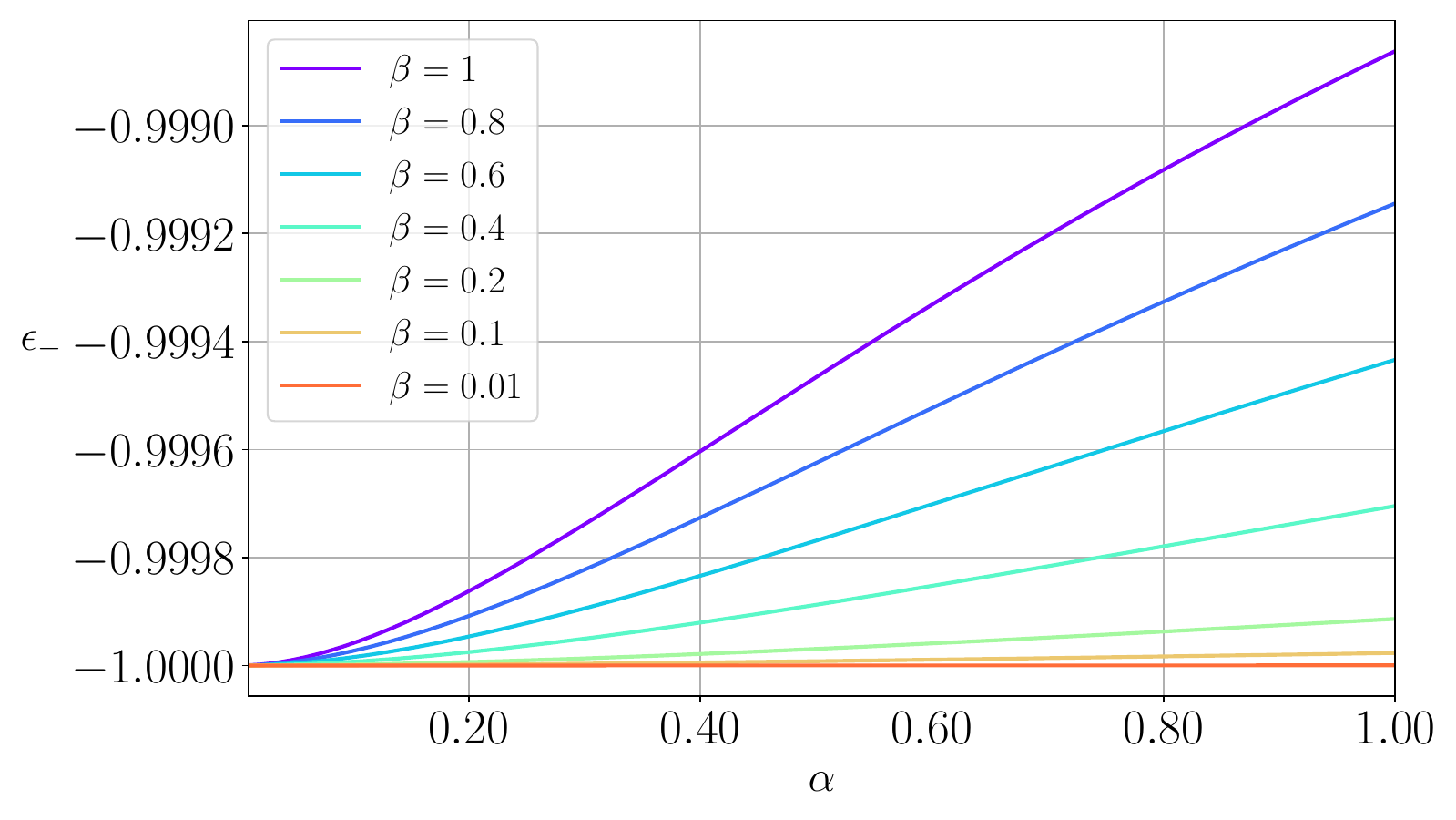}
    \caption{Plots of energy against angular deficit for the $N=0$ energy level for different values of $\beta$, with parameters $\delta_r = 0.5$, $\delta_t = 0.6$, $\gamma_s = 0.7$, $e\gamma_t = -0.8$, $\alpha = 0.9$, $m = 1$, and $M=1$.}
    \label{fig:EnergyLevelsMinusVsAlpha-SeveralBeta}
\end{figure}

\textcolor{black}{Continuing with the examination of the global monopole parameter's impact, as illustrated in \autoref{fig:EnergyLevelsMinusVsAlpha-SeveralBeta}, for $\beta=1$, the anticipated outcome as found by \cite{neto2020scalar} is confirmed. Furthermore, when the symmetry-breaking scale associated with monopole formation is increased, $\beta$ decreases, and as $\beta$ approaches zero, the energy level is seen to converge towards the asymptotic value $-M$.}

\begin{figure}[H]
    \centering
    \includegraphics[scale=0.5]{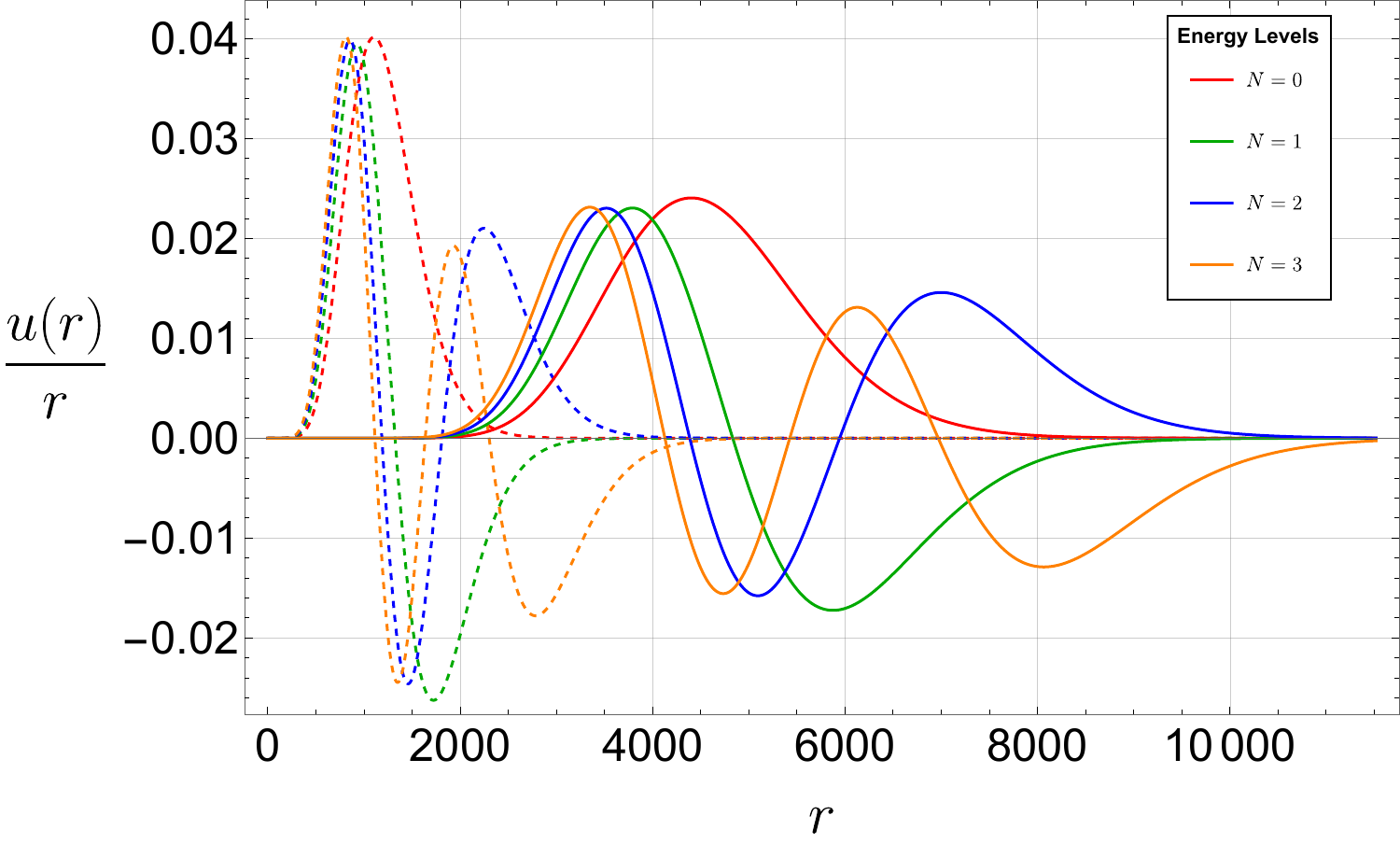}
    \caption{Plot of the radial wave functions for four different values of $N$, for both $\beta=1$ (dashed lines) and $\beta=0.5$ (solid lines) with parameters $\delta_r = 0.5$, $\delta_t = 0.6$, $\gamma_s = 0.7$, $e\gamma_t = -0.8$, $\alpha = 0.1$, $m = 1$, $l=1$, and $M=1$.}
    \label{fig:NormalizedRadialWaveFunctionBetaDashedSolid}
\end{figure}

\begin{figure}[H]
    \centering
    \includegraphics[scale=0.5]{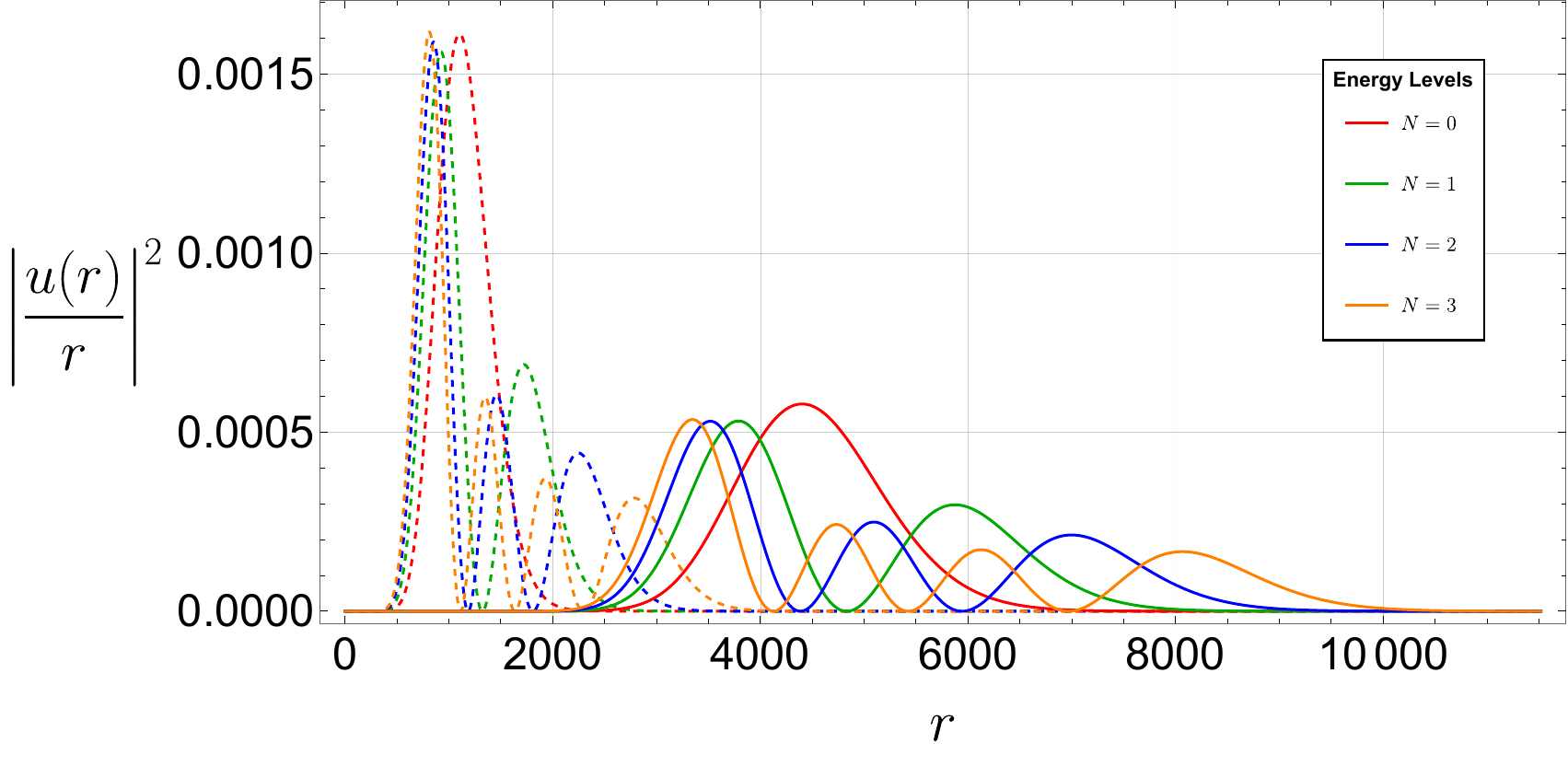}
    \caption{Plot of the radial probability density for four different values of $N$, for both $\beta=1$ (dashed lines) and $\beta=0.5$ (solid lines) with parameters $\delta_r = 0.5$, $\delta_t = 0.6$, $\gamma_s = 0.7$, $e\gamma_t = -0.8$, $\alpha = 0.1$, $m = 1$, $l=1$, and $M=1$.}
    \label{fig:NormalizedRadialWaveFunctionSquaredBetaDashedSolid}
\end{figure}

As an additional condition, note that $\varepsilon$ is real for $e^2 \gamma_t^2 > \gamma_s^2$. Equation (\ref{hiper}) can be written in the form
\begin{align}
    u(r) &= Ae^{-|K|r}(2|K|r)^{1/2 + \gamma_l}M(-N,1+2\gamma_l,2|K|r)\nonumber\\
    & = N_n e^{-|K|r} (2|K|r)^{1/2 + \gamma_l}L^{\:2\gamma_l}_{N}(2|K|r), \label{eq:laguerrewavefunc}
\end{align}
where $L_{N}^{\:a-1}$ is the generalized Laguerre polynomial and $N_n$ is the normalization constant. \textcolor{black}{Figures \ref{fig:NormalizedRadialWaveFunctionBetaDashedSolid} and \ref{fig:NormalizedRadialWaveFunctionSquaredBetaDashedSolid} illustrate the radial wave function and probability densities from Eq. \ref{eq:laguerrewavefunc} for the initial four $N$ values, given specific parameters for the scenarios without a global monopole ($\beta=1$) and for $\beta=0.5$. As can be seen, the plots reveal that as the symmetry-breaking scale associated with monopole formation increases, the wave function becomes more spread out. Consequently, the probability density exhibits a greater dispersion along the radial coordinate.}

\textcolor{black}{Just like we did for the scattering states, an investigating of the non-relativistic limit of the energy spectrum can be done. Considering now the approximation (\ref{Approx}) in the equation that characterizes the energy spectrum (\ref{Eq_Espect}), and retaining only first--order terms in $E_{\text{nr}}$, one can write
\begin{equation}
  E_{\text{nr}}=-\frac{M\chi^{2}}{2\left[\chi e\gamma_{t}+\left(N+\frac{1}{2}+\sqrt{\frac{1}{4}+\frac{l_{\alpha}\left(l_{\alpha}+1\right)}{\beta^{2}}}\right)^{2}\right]},
\end{equation}
where $\chi=e\gamma_{t}-\gamma_{s}$, this expression explicitly depends on the topological defects $\alpha$ and $\beta$, and on the parameters $\gamma_t$ and $\gamma_s$. Setting the vector coupling to zero ($\gamma_t=0$) yields
\begin{equation}
    E_{\text{nr}} = -\frac{M\gamma_{s}^{2}}{2\left(N+\frac{1}{2}+\sqrt{\frac{1}{4}+\frac{l_{\alpha}\left(l_{\alpha}+1\right)}{\beta^{2}}}\right)^{2}}.
\end{equation}}

\textcolor{black}{Furthermore, by removing the topological defect associated with the cosmic string ($\alpha=1$), the result obtained in \cite{BezerradeMello:1996si} is recovered:
\begin{equation}
    E_{\text{nr}} = -\frac{M\gamma_{s}^{2}}{2\left(N+\frac{1}{2}+\sqrt{\frac{1}{4}+\frac{l\left(l+1\right)}{\beta^{2}}}\right)^{2}}.
\end{equation}}

\textcolor{black}{Finally, by removing the global monopole defect ($\beta=1$), the energy spectrum in flat space is obtained:
\begin{equation}
    E_{\text{nr}} = -\frac{M\gamma_{s}^{2}}{2\left(N+l+1\right)^{2}}.
\end{equation}
In other words, the appropriate limits for the topological defects and the Coulomb potential coupling parameters recover well--known solutions in the literature, which indicates the consistency of the generalizations introduced.}

\section{Particular Cases} \label{sec5}

At this point, it is worth studying the special cases that can be obtained from the particular values of the potential parameters. Naturally, the conditions for obtaining bound states indicate that we must analyze the bound states separately for each particular solution. Note that the pure nonminimal vector potential $X_{\mu}$ does not provide bound states and is therefore not considered. 

\subsection{Minimally coupled vector potential}
In this case, we consider only the electromagnetic vector potential, which implies that $\delta_t =\delta_r =\gamma_s=0$. The condition $\alpha_1 < 0$ demands that bound states are possible only if $2e\varepsilon \gamma_t > 0$. Thus, we have two conditions, depending on the signal of $\gamma_t$ and $\varepsilon$ that satisfy this requirement. The energy spectrum for this particular case reduces to 
\begin{equation}
    \varepsilon_+ = \frac{M}{\sqrt{1+\frac{e\gamma^2_t}{\mu^2}}},\text{\:\:if\:\:} \gamma_t > 0, 
    \label{eq5.1}
\end{equation}
or
\begin{equation}
    \varepsilon_- = -\frac{M}{\sqrt{1+\frac{e\gamma^2_t}{\mu^2}}},\text{\:\:if\:\:} \gamma_t < 0, 
    \label{eq5.2}
\end{equation}
where 
\begin{align}
  \mu &= N + 1/2 + \gamma_l \nonumber\\
      & =N + 1/2 + \sqrt{\frac{1}{4} + \frac{l_\alpha(l_\alpha+1)}{\beta^2} -e^2\gamma^2_t} \quad.
\end{align}

For $\beta = 1$, the energy spectrum obtained reduces to the Coulomb potential in the cosmic string spacetime \cite{neto2020scalar}, while the case $\alpha = \beta = 1$ corresponds to the solution of the Klein-Gordon equation with a Coulomb potential in a flat geometry \cite{greiner2000relativistic}. \textcolor{black}{From a physical point of view, the particular solution studied here can be used to describe the behavior of charged particles under the effect of a Coulomb potential in a spacetime with topological defect. As we can see from Equations (\ref{eq5.1}) and (\ref{eq5.2}), the interaction between the Coulomb potential and the conical geometry modifies the energy spectrum of the scalar particle. Another feature of this solution is the presence of particles and antiparticles.}
\subsection{Scalar potential}
Let us consider the particular case where $\delta_t =\delta_r =\gamma_t=0$. Taking into account the condition $\alpha_1 < 0$, bound states are possible only if $2M\gamma_s < 0$. As a result, Eq. (\ref{energia}) reduces to
\begin{equation}
    \varepsilon_{\pm} = \pm M \sqrt{1-\frac{\gamma^2_s}{\mu^2}} \quad,
\end{equation}
where 
\begin{align}
  \mu &= N + 1/2 + \gamma_l \nonumber\\
      & =N + 1/2 + \sqrt{\frac{1}{4} + \frac{l_\alpha(l_\alpha+1)}{\beta^2} +\gamma^2_s} \quad.
\end{align}

If $\beta = 1$, the spectrum obtained reduces to the scalar Coulomb potential in the cosmic string spacetime \cite{neto2020scalar}. The case $\alpha = \beta = 1$ is associated with a solution of the Klein-Gordon equation with a scalar Coulomb potential in a flat geometry 
\cite{greiner2000relativistic}. \textcolor{black}{The scalar potential couples to the mass of the particle, so that we can see the emergence of an effective mass for the scalar particle. In contrast to the vector potential that couples to the four-current of the particle, a scalar potential does not distinguish particles from antiparticles. This mechanism created by the scalar potential has general applications in nuclear and particle physics, in particular, it can simulate effective mass in solids \cite{neto2020scalar}. }
\subsection{Mixed scalar-vector potentials. }
Now, we have the conditions $\gamma_s/r \neq0$ and $\gamma_t/r \neq 0$ while $\delta_r/r = \delta_r/r = 0$.  The expression for bound state $\alpha_1 < 0$ requires that the parameters $M,\varepsilon,\gamma_t$ and $\gamma_s$ are connected in the form $2(M-e\varepsilon\gamma_t) < 0$. The final form of expression (\ref{energia}) is given by 
\begin{equation}
    \varepsilon_{\pm} = \frac{M}{\left(1 + \frac{e^{2}\gamma_{t}^{2}}{\mu^{2}}\right)} \left( \frac{\gamma_{s}}{\mu} \frac{e\gamma_{t}}{\mu} \pm \sqrt{1 + \frac{e^{2}\gamma_{t}^{2}}{\mu^{2}} - \frac{\gamma_{s}^{2}}{\mu^{2}}} \right)
    \label{energia2},
\end{equation}
where
\begin{align}
  \mu &= N + 1/2 + \gamma_l \nonumber\\
      & =N + 1/2 + \sqrt{\frac{1}{4} + \frac{l_\alpha(l_\alpha+1)}{\beta^2} +\gamma^2_s-e^2\gamma_t^2} \quad.
      \label{mixed}
\end{align}
 
By considering $\beta = 1$, we obtain the same expression for the energies for the Klein-Gordon equation with both scalar and vector potentials in the cosmic string spacetime \cite{neto2020scalar}. In flat spacetime case, $\beta = 1$ and $\alpha = 1$, the energy spectrum agrees with the literature \cite{garcia2015relativistic}. \textcolor{black}{Physically, this solution describes a charged particle under the effects of two distinct interactions. It can be seen as a combination of the previous particular solutions. The vector Coulomb potential leads to an interaction that mimics the hydrogen atom but with relativistic corrections. The scalar Coulomb potential modifies the mass term of the Klein-Gordon equation, leading to stronger binding energies and different spectral properties. Since it does not couple to charge, it treats particles and antiparticles symmetrically.}

\section{Discussion and Conclusions} \label{sec6}

This work explored the effects of cosmic strings and global monopoles on the quantum dynamics of bosonic particles through a generalized metric that accounts for both defects. We considered a metric that reflects the influence of the topological features, with the parameters \(\alpha\) and \(\beta\) characterizing the deficit angles introduced by the cosmic string and global monopole, respectively. By applying this metric to the Klein-Gordon equation with generalized Coulomb potentials, we investigated how scalar bosons behave under the influence of these potentials and topological defects. The equation included scalar and minimal vector potentials, alongside nonminimal couplings, capturing the effects of both external fields and spacetime geometry. Solutions for the wave function were obtained, revealing that both the angular and radial components are modified by the topological parameters.

{\color{black} The solution of the wave function exhibits an angular dependence governed by Eq. \ref{3.6}, which is modified by the parameter $\alpha$, while its radial behavior follows Eq. \ref{3.8} and is influenced by $\beta$, as illustrated in Figures \ref{fig:NormalizedRadialWaveFunctionBetaDashedSolid} and \ref{fig:NormalizedRadialWaveFunctionSquaredBetaDashedSolid}, demonstrating that decreasing $\beta$ results in a delocalization of the radial probability density. The bound state energy spectrum is given by Eq. \ref{energia}, and its dependence on the principal quantum number $N$ and the angular momentum quantum number $l$ can be visualized in Fig. \ref{fig:EnergySpectraxNxl-Beta0.6}, whereas Fig. \ref{fig:EnergyLevelsMinusVsAlpha-SeveralBeta} highlights how the energy $\varepsilon_{\pm}$ approaches $\pm M$ as the global monopole parameter decreases. Additionally, considering limiting cases provides further insight into the underlying physical implications: setting $\beta = 1$ recovers the well-known cosmic string results \cite{neto2020scalar}, while taking $\alpha = \beta = 1$ reduces to the flat spacetime result \cite{greiner2000relativistic}. This correspondence is reinforced by Figures \ref{fig:EnergySpectraBeta0.8And1} and \ref{fig:EnergyLevelsMinusVsAlphaBeta0.3And1}, which demonstrate how these cases emerge as particular limits of the general solution. Moreover, Fig. \ref{fig:energyAgainstBeta} reveals that states with higher values of $l$ exhibit a more pronounced decrease in $\varepsilon_{-}$, further emphasizing the role of angular momentum in shaping the energy spectrum.
}

In particular, the study provided findings in the study of bound and scattering states under the combined influence of scalar ($V_s$) and vector  potentials ($A_\mu$ and $X_\mu$). We obtain an effective radial equation that resembles a Schrödinger-type differential equation with a Coulomb-like potential, allowing us to explore both bound and scattering states solutions. We showed that bound states are determined by specific conditions related to the defect parameters, such as the deficit angles and the constants characterizing the potentials. Additionally, particular cases were analyzed, including pure scalar and vector potentials, as well as mixed scalar-vector potentials. In particular, we consider bound states where $\alpha_1 <0$ in addition to the condition for the energy $|M| < \varepsilon$. As we can see from particular cases, these expressions impose different relations between the parameters of potentials and bosons. In the case of minimally coupled vector potential, depending on values for $\gamma_t$. Positive/negative energy levels are related to positive/negative values for $\gamma_t$. For the pure scalar potential case, the equation $2M\gamma_s <0$ imposes that the only possible values obey $\gamma_s < 0$. The energy spectrum for the Klein-Gordon equation with mixed scalar-vector potentials is given by expression (\ref{mixed}) where we can see that there is a coupling term between the parameters of the potentials $\gamma_s$ and $\gamma_t$. We do not consider pure nonminimal vector potentials due to the fact that the parameters of this case do not lead to bound states solutions.  The study of these particular cases demonstrates the possibility of several physical systems being described by a Klein-Gordon equation with generalized potentials where the energy spectra reduce to well-known solutions in the absence of defects or certain interactions. 

\textcolor{black}{Furthermore, in the non-relativistic limit, the phase shift and energy spectrum exhibit expected behaviors consistent with known results. When the cosmic string defect is absent ($\alpha=1$), the phase shift and energy levels coincide with those found in a global monopole background. Similarly, setting $\beta=1$ eliminates the effects of the global monopole, recovering the standard Coulomb expressions in cosmic string spacetime. In the combined limit ($\alpha=\beta=1$), all modifications due to topological defects vanish, and the system fully reproduces the conventional non-relativistic Coulomb problem. These limiting cases further validate the consistency of the generalized formulation.
}

Our results are summarized below:
\begin{itemize}
    \item Use of a generalized metric that incorporates the effects of both cosmic strings and global monopoles, with the parameters \(\alpha\) and \(\beta\) representing their associated deficit angles.    
    \item Inclusion of scalar, minimal vector, and nonminimal vector coupling terms in the Klein-Gordon equation, capturing the combined effects of spacetime geometry and external potentials.  
    \item Solution of the wave function, showing that both angular and radial components are modified by the topological parameters \(\alpha\) and \(\beta\).  
    \item Derivation of an effective radial equation resembling a Schrödinger-type differential equation with a Coulomb-like potential, enabling the study of bound and scattering state solutions.  
    \item Identification of the specific conditions required for bound states, including constraints on the defect parameters, such as deficit angles, and on the potential constants.  
    \item Analysis of particular cases, including pure scalar, pure vector, and mixed scalar-vector potentials, with expressions that establish relations between the potential parameters and bosonic energies.  
    \item Exploration of the minimally coupled vector potential case, showing that positive and negative energy levels are related to the sign of the parameter \(\gamma_t\).  
    \item Study of the pure scalar potential case, revealing that bound states are only possible for \(\gamma_s < 0\) under the condition \(2M\gamma_s < 0\).  
    \item Investigation of mixed scalar-vector potentials, demonstrating how the coupling between the parameters \(\gamma_s\) and \(\gamma_t\) affects the energy spectrum.  
    \item Demonstration that the Klein-Gordon equation with generalized potentials can describe various physical systems, with energy spectra reducing to known solutions in the absence of topological defects or specific interactions.    
\end{itemize}

Overall, the results highlight how the presence of topological defects can affect the dynamics of quantum particles. Future studies could further explore these effects in different contexts, such as fermionic fields or other forms of couplings, to gain better knowledge about the influence of defects on quantum systems.

\begin{acknowledgements}
LGB and JVZ acknowledge the financial support from CAPES (process numbers 88887.968290/2024-00 and 88887.655373/2021-00, respectively). LCNS would like to thank FAPESC for financial support under grant 735/2024.
\end{acknowledgements}

\bibliographystyle{spphys}
\bibliography{referencias_unificadas2}
\end{document}